\documentclass[aps,superscriptaddress,amsmath,amssymb,floatfix,twocolumn,showpacs,amsfonts]{revtex4-1}
\usepackage{times}
\usepackage[varg]{txfonts}
\usepackage{textcomp}
\usepackage{graphicx}
\usepackage{color}
\usepackage[colorlinks=true,citecolor=blue,urlcolor=blue,linkcolor=blue,hyperindex]{hyperref}
\usepackage{braket}
\usepackage{overpic}
\usepackage{multirow}
\usepackage[table ]{ xcolor}

\def\kk{\mathbf{k}}
\def\qq{\mathbf{q}}
\def\rr{\mathbf{r}}
\def\dd{\mathbf{d}}
\def\uu{\mathbf{u}}
\def\vv{\mathbf{v}}
\def\bs{\mathbf{S}}

\def\bq{\mathbf{Q}}

\def\half{\frac{1}{2}}

\begin{document}

\title{Spin and quadrupolar orders in the spin-1 bilinear-biquadratic model for iron-based superconductors}

\author{Cheng Luo}
\affiliation{State Key Laboratory of Optoelectronic Materials and Technologies, School of Physics, Sun Yat-Sen University, Guangzhou 510275, China}

\author{Trinanjan Datta}
\affiliation{Department of Chemistry and Physics, Augusta University, 1120 15th Street, Augusta, Georgia 30912, USA}

\author{Dao-Xin Yao}
\email[Corresponding author:]{yaodaox@mail.sysu.edu.cn}
\affiliation{State Key Laboratory of Optoelectronic Materials and Technologies, School of Physics, Sun Yat-Sen University, Guangzhou 510275, China}

\date{\today}

\begin{abstract}
Motivated by the recent experimental and theoretical progress of the magnetic properties in iron-based superconductors,
we provide a comprehensive analysis of the extended spin-1 bilinear-biquadratic (BBQ) model on the square lattice.
Using a variational approach at the mean-field level, we identify the existence of various magnetic phases,
including conventional spin dipolar orders (ferro- and antiferromagnet), novel quadrupolar orders (spin nematic)
and mixed dipolar-quadrupolar orders. In contrast to the regular Heisenberg model,
the elementary excitations of the spin-1 BBQ model are described by the SU(3) flavor-wave theory.
By fitting the experimental spin-wave dispersion,
we determine the refined exchange couplings corresponding to the collinear antiferromagnetic iron pnictides.
We also present the dynamic structure factors of both spin dipolar
and quadrupolar components with connections to the future experiments.

\begin{description}
\item[PACS number(s)] 74.70.Xa,75.10.-b,75.25.-j,75.10.Jm
\end{description}
\end{abstract}

\maketitle

%%%%%%%%%%%%%%%%%%%%%%%%%%%%%%%%%%%%%%%%%%%%%%%%%%%%%%%%%%%%%%%%
\section{Introduction}
%%%%%%%%%%%%%%%%%%%%%%%%%%%%%%%%%%%%%%%%%%%%%%%%%%%%%%%%%%%%%%%%

The parent compounds of the recently discovered iron-based superconductors (SCs)
exhibit a variety of unusual magnetic phases, reviewed by Ref.~\cite{RevModPhys.87.855} and references therein.
Instead of the regular N\'{e}el antiferromagnetic (AFM) order found in cuprates,
iron pnictides display a collinear antiferromagnetic (CAFM) order.
In contrast to cuprates superconductors where the magnetism of the parent compounds are well described by
a nearest-neighbor (NN) Heisenberg model, the character of magnetic interactions in iron-based SCs may not be well described by Heisenberg-type models.

Early inelastic neutron scattering (INS) experiments in the iron pnictides reveal that the spin-wave excitations in
these compounds are highly anisotropic~\cite{NatPhys.5.555,PhysRevB.83.214519,PhysRevB.84.054544},
with a dispersion which can be understood in terms of a phenomenological $J_x-J_y-J_2$ model.
However, the strong anisotropic version with antiferromagnetic $J_x$ and ferromagnetic $J_y$
is not compatible with the tetragonal lattice structure even when the small orthorhombic distortion is
taken into account. Thus a correct interpretation of the strong anisotropy entails
additional underlying mechanisms.
While the strong anisotropy of magnetic interactions in iron-based SCs has also been linked to nematic ordering~\cite{NatPhys.10.97},
there is still an ongoing debate on whether it is caused by the spin-nematic sector~\cite{PhysRevB.77.224509,PhysRevB.78.020501,Dai17032009,PhysRevLett.107.217002}
or the orbital ordering sector~\cite{PhysRevB.79.054504,PhysRevB.80.224506,PhysRevLett.103.267001,PhysRevB.80.180418,PhysRevB.82.045125}.

Based on a simple local moment picture, this puzzling feature found experimentally can be naturally explained
with the inclusion of a biquadratic spin coupling.
Indeed, the parent compounds of a large majority of iron-based SCs host a range of semimetallic behaviors,
signaling the deviation of Mott insulating state,
it is natural that the magnetic Hamiltonian consists of not only bilinear spin interactions originating from
the strong coupling regime but also interactions involving multi-spin exchange terms
when perturbation expansion is carried out up to fourth order~\cite{Fazekas1999,EurPhysJB}.
The presence of biquadratic terms in iron-based materials has been verified
by first principle calculations~\cite{PhysRevB.79.144421,PhysRevB.89.064509}.
It is argued that the biquadratic spin-spin interaction describes the low-energy properties after
integrating out the itinerant electrons or orbital degrees of freedom~\cite{NatPhys.7.485,JPCS012024}.
Furthermore, a large biquadratic exchange in iron-based SCs was also attributed to the crossover of
different local spin states~\cite{PhysRevLett.110.207205}.

In the previous studies of the CAFM iron pnictides, only the NN biquadratic coupling $K_1$ was included
which leads to a minimum effective $J_1-J_2-K_1$ model
~\cite{NatPhys.7.485,PhysRevB.84.064505,PhysRevB.85.144403,PhysRevB.86.085148,PhysRevB.92.165102}.
This minimum model can really preserve the tetragonal lattice symmetry
and capture the essentially anisotropic spin excitations of the parent compounds CaFe$_2$As$_2$ and BaFe$_2$As$_2$.
However, it is not a prior that the next NN (NNN) biquadratic coupling $K_2$ is negligible,
since both experimental and theoretical studies suggest substantial superexchange process though NNN sites.
Therefore we consider the extended bilinear-biquadratic (BBQ) model defined as
\begin{align}\label{BBQmodel}
 \mathcal{H}=\sum_{\mu=1,2}\sum_{\langle ij\rangle_\mu}J_\mu\bs_i\cdot\bs_j-K_\mu(\bs_i\cdot\bs_j)^2,
\end{align}
where $\langle ij\rangle_1$ and  $\langle ij\rangle_2$ denote the NN and NNN bonds, respectively.
We note that we have not considered the third NN couplings which are believed to be essential to describe the
magnetic properties of iron chalcogenides~\cite{PhysRevB.85.144403,PhysRevLett.115.116401},
since the above model appears to be adequate to describe iron pnictides.

In the present work, we assume the effective spin $S=1$ on the iron sites
based on the successful studies of two-band models~\cite{PhysRevB.77.220503}. The reason is two-fold,
(i) $S\leq1$ agrees with the observed relatively small local moments from the integrated spin spectral
weight of INS measurements, (ii) the biquadratic spin interactions are expected to be a natural consequence of the
strong coupling expansion in multi-orbital systems with local effective spin $S\geq1$ induced by Hund's coupling.
The later can be easily understood by recasting the biquadratic term as
\begin{align}\label{BQterm}
 2(\bs_i\cdot\bs_j)^2=\bq_i\cdot\bq_j-\bs_i\cdot\bs_j+\frac{2}{3}S^2(S+1)^2,
\end{align}
where $\bq$ is the quadrupolar operator with five components
$Q^{\alpha\beta}=\frac{1}{2}\{S^\alpha,S^\beta\}-\frac{1}{3}S(S+1)\delta_{\alpha\beta}$~\cite{Penc2010}.
It is worth noting that $\bq=0$ for $S=\half$ and the BBQ model (\ref{BBQmodel}) reduces to
an effective $\bar{J_1}-\bar{J_2}$ model with renormalized exchange constants
$\bar{J_1}=J_1+\frac{K_1}{2}$ and $\bar{J_2}=J_2+\frac{K_2}{2}$.
Though it was shown that biquadratic coupling could be generated
by quantum or thermal fluctuations in a bare $J_1-J_2$ model~\cite{PhysRevLett.64.88,PhysRevB.77.224509},
the small amplitude of the biquadratic constant is not applicable to experiments.
Thus it is commonly believed that the Heisenberg
$J_1-J_2$ model can not explain the observed anisotropic spin
excitations in iron pnictides, this justifies the validity of using $S=1$ for our study.

In this paper, we study the spin-1 BBQ model by treating the spin dipolar and quadrupolar
degree of freedom on an equal footing.
Despite conventional spin dipolar orders with finite magnetic moments $\langle\bs\rangle$
found previously~\cite{PhysRevB.85.144403}, our variational phase diagram by taking into account
the quantum nature of local spin-1 states show that the BBQ model can support novel orderings of
spin quadrupolar moments $\langle\bq\rangle$ which are also known as spin nematic phases without
time-reversal symmetry breaking~\cite{PhysRevLett.97.087205,TsunetsuguJPSJ75083701,PhysRevLett.96.027213}.
Contrary to previous works based on  conventional SU(2) spin-wave theory where the intrinsic quadrupolar fluctuations
were missed~\cite{NatPhys.7.485,PhysRevB.84.064505,PhysRevB.85.144403,PhysRevB.86.085148,PhysRevB.92.165102},
we find that the elementary excitations in the framework of SU(3) flavor-wave theory
display distinct features which are crucial for a consistent interpretation of the
magnetic interactions and spin excitation spectra in iron-based superconductors.

The remainder of the paper is organized as follows. In Sec.~\ref{Sec.ground state},
we identify various variational ground states of the BBQ model on the square lattice and
display a portion of the mean-field phase diagram relevant for the iron-based SCs.
In Sec.~\ref{Sec.flavor wave}, we introduce the SU(3) flavor-wave theory for spin-1 systems
and provide its implications for the spin dynamics of CAFM iron pnictides.
The comparison with the results given by conventional SU(2) spin-wave theory is also made.
In Sec.~\ref{Sec.dynamic correlations} we present the dynamic correlation functions of
both spin dipolar and quadrupolar components with connections to the future experiments in iron pnictides.
Sec.~\ref{Sec.conclusion} is devoted to our summary and conclusion.
For the sake of completeness, we present the formula of the conventional SU(2) spin-wave theory
in the Appendix.

%%%%%%%%%%%%%%%%%%%%%%%%%%%%%%%%%%%%%%%%%%%%%%%%%%%%%%%%%%%%%%%%
\section{Variational ground states}\label{Sec.ground state}
%%%%%%%%%%%%%%%%%%%%%%%%%%%%%%%%%%%%%%%%%%%%%%%%%%%%%%%%%%%%%%%%

\subsection{Parametrization of spin-1 states}
We discuss the zero temperature variational phase diagram
based on the following site-factorized wave function~\cite{Penc2010,PhysRevLett.97.087205}
\begin{align}
 \ket{\Psi}=\prod_{i=1}^N\ket{\psi_i},
\end{align}
where $N$ is the number of lattice sites and $\ket{\psi_i}$ the local wave function at site $i$.
It is convenient to introduce the time-reversal invariant basis for spin-1 states
\begin{align}
 \ket{x}=i\frac{\ket{1}-\ket{\bar{1}}}{\sqrt{2}},\ \ket{y}=\frac{\ket{1}+\ket{\bar{1}}}{\sqrt{2}},\ \ket{z}=-i\ket{0},
\end{align}
where $\ket{\bar{1}},\ket{0},\ket{1}$ are the usual bases quantized along the $z$ axis.
A general single-site wave function can be written as
\begin{align}
  \ket{\psi_i}=\sum_{\alpha=x,y,z} d_{i\alpha}\ket{\alpha},
\end{align}
where $\dd=\uu+i\vv$ satisfies the normalization constraint $|\dd|=1$.
Without loss of generality, one can choose $\uu$ and $\vv$ in such a way that $\uu\cdot\vv=0$.
The coherent spin state is realized for $u=v$ and $\langle\bs\rangle=2\uu\times\vv$.
If $u=0$ or $v=0$, the state is purely quadrupolar with a \textit{director} along the nonzero component $\uu$ or $\vv$.
In fact, one may refer to the larger of the two vectors as the director in the case $0<\langle\bs\rangle<1$
(in other words, the spin is not fully developed).
It is evident that the director has to lie in the plane perpendicular to the spin vector in the partially developed state.

\subsection{Large-$J$ ground state}

Using Eq.~(\ref{BQterm}), we can recast the BBQ Hamiltonian (\ref{BBQmodel}) as,
up to a constant energy ($-\frac{8}{3}N(K_1+K_2))$
\begin{align}\label{BBQmodel1}
 \mathcal{H}=\sum_{\mu=1,2}\sum_{\langle ij\rangle_\mu}J^D_\mu\bs_i\cdot\bs_j
 -J^Q_\mu\bq_i\cdot\bq_j,
\end{align}
where we have defined the effective spin dipolar and quadrupolar couplings $J^D_\mu=J_\mu+\frac{K_\mu}{2}$ and
$J^Q_\mu=\frac{K_\mu}{2}$, respectively.
In order to provide a simple but rather instructive picture,
the minimization of the mean-field ground state energy $E_{\mathrm{GS}}=\bra{\Psi}\mathcal{H}\ket{\Psi}$
can be carried out first for several extreme cases with only one type of coupling surviving, which we term large-$J$ limit.
Since the exchange couplings involve both NN and NNN bonds, one obtains eight different
ground state manifolds depending on the sign of the coupling coefficient.
Some of these ground state manifolds are unique while others are degenerate.
The variational results are summarized in Table~\ref{Table:Groundstate}.
We will present the detailed analysis for each case in the following.
%%%%%%%%%%%%%%%%%%%%%%%%%%%%%%%%
\begin{table}
\caption{\label{Table:Groundstate}
Variational ground state manifolds under individual interaction $J_\mu^D$ or $J_\mu^Q$ on the square lattice.
}
\begin{ruledtabular}
\begin{tabular}{cccc}
Interaction            &  Sign & Ground state & Degeneracy\\ \hline
\multirow{2}{*}{$J^D_1$} &   +   & (N\'{e}el) AFM  & 1\\
                       &   -   & FM   & 1\\ \hline
\multirow{2}{*}{$J^Q_1$} &   +   & FQ  & 1\\
                       &   -   & Semi-ordered (SO)   & $\infty$\\ \hline
\multirow{2}{*}{$J^D_2$} &   +   & Decoupled AFM  & $\infty$\\
                       &   -   & Decoupled FM   & $\infty$\\ \hline
\multirow{2}{*}{$J^Q_2$} &   +   & Decoupled FQ  & $\infty$\\
                       &   -   & Decoupled SO   & $\infty$\\
\end{tabular}
\end{ruledtabular}
\end{table}
%%%%%%%%%%%%%%%%%%%%%%%%%%%%%%%%

(i) Large-$J^D_1$ limit. We may set $|J^D_1|=1$ and $J^Q_1=J^D_2=J^Q_2=0$.
In this case we find only spin dipolar operators among NN bonds are coupled.
The presence of only $J^D_1$ interaction induces conventional ferromagnetic (FM) phase for $J^D_1<0$
and N\'{e}el AFM phase for $J^D_1>0$.

\begin{figure}
\centering\includegraphics[scale=0.5]{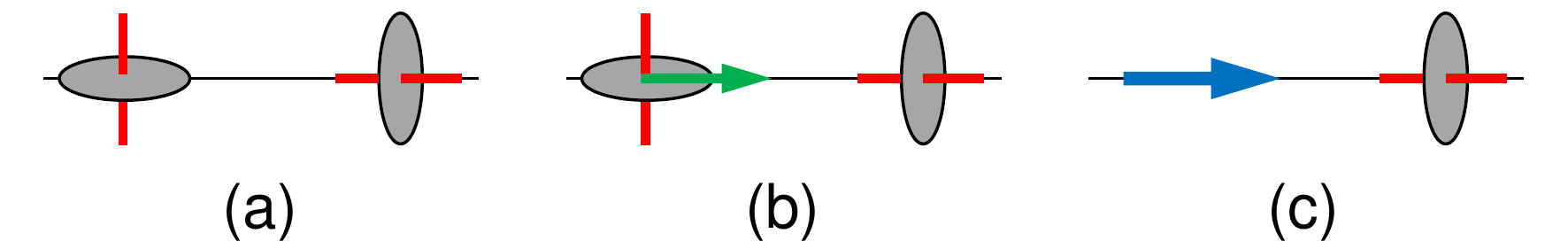}
\caption{(Color online)
Degeneracy of a semi-ordered (SO) bond with dominant AFQ coupling among two spin-1 sites.
Keeping one of the sites a purely quadrupoalr state, one may either obtain the solution
of (a) a quadrupole , (b) a partially developed dipole or (c) a coherent spin state on the other site.
Blue (green) arrows symbolize coherent (partially developed) moments, while red lines symbolize directors.
}
\label{fig:SObond}
\end{figure}

(ii) Large-$J^Q_1$ limit. We may set $|J^Q_1|=1$ and $J^D_1=J^D_2=J^Q_2=0$. In this case
we find only spin quadrupolar operators among NN bonds are coupled.
It is easy to show that a ferroquadrupolar (FQ) phase with parallel directors for all the sites
will be stabilized for $J^Q_1>0$. However, the case for antiferroquadrupolar (AFQ) coupling $J^Q_1<0$ is nontrivial.
In order to gain further insight, one should note that the expectation value of a pair of
quadrupolar operators can be written as
\begin{align}
 \langle\bq_i\cdot\bq_j\rangle=|\dd_i\cdot\dd_j|^2+|\dd_i^\ast\cdot\dd_j|^2-\frac{2}{3}.
\end{align}
We see that $\langle\bq_i\cdot\bq_j\rangle$ is minimized if $\dd_j$ is orthogonal both to $\dd_i$ and its time-reversal transform,
which implies that one state is a pure quadrupole with director $\dd$ while another one may feature either a pure
quadrupole with its director perpendicular to $\dd$, or a spin vector of arbitrary length pointing along $\dd$.
Following Ref.~\cite{Papanicolaou1988367,PhysRevLett.105.265301,PhysRevB.85.140403},
we call this phase "semi-ordered" (SO).
The degeneracy of a SO bond with dominant AFQ coupling among two sites is depicted in Fig.~\ref{fig:SObond}.
%%%%%%%%%%%%%%%%%%%%%%%%%%%%%%%%
%%%%%%%%%%%%%%%%%%%%%%%%%%%%%%%%

\begin{figure}
\centering\includegraphics[scale=0.38]{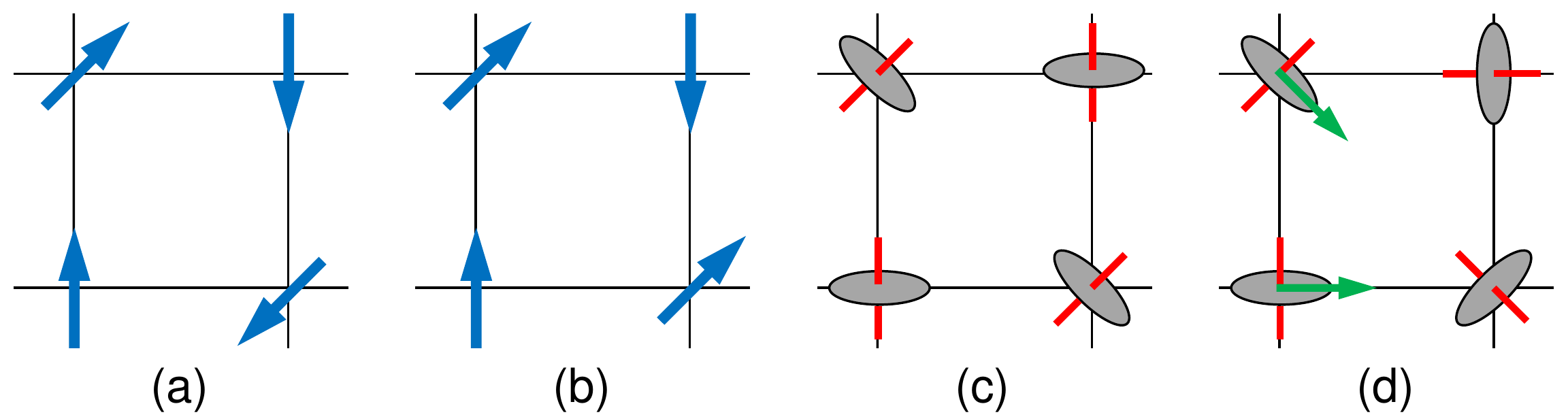}
\caption{(Color online)
Illustration of the four decoupled phases in the presence of individual NNN interaction $J_2^D$ or $J_2^Q$ on a plaquette.
(a) Decoupled AFM for $J_2^D>0$. (b) Decoupled FM for $J_2^D<0$.
(c) Decoupled FQ for $J_2^Q>0$. (d) Decoupled SO for $J_2^Q<0$.
}
\label{fig:Decoupled}
\end{figure}

(iii) Large-$J^D_2$ limit. We may set $|J^D_2|=1$ and $J^D_1=J^Q_1=J^Q_2=0$.
In this case we find only spin dipolar operators among NNN bonds are coupled.
In the presence of only $J^D_2$ interaction, the lattice decouples into two interpenetrating FM ($J^D_2<0$)
or AFM ($J^D_2>0$) sublattices and the angle between the magnetization or staggered magnetization
of these two sublattices is arbitrary.
The decoupled AFM and FM phases on a plaquette are depicted in Fig.~\ref{fig:Decoupled} (a) and (b), respectively.
%%%%%%%%%%%%%%%%%%%%%%%%%%%%%%%%
%%%%%%%%%%%%%%%%%%%%%%%%%%%%%%%%

(iv) Large-$J^Q_2$ limit.  We may set $|J^Q_2|=1$ and $J^D_1=J^Q_1=J^D_2=0$.
In this case, we find only spin quadrupolar operators among NNN bonds are coupled.
In analogy with the above analysis, we also find two decoupled phases with decoupled FQ for $J^Q_2>0$
and decoupled SO for $J^Q_2<0$.
We depict two examples on a plaquette in Fig.~\ref{fig:Decoupled} (c) and (d).

\subsection{Lift of degeneracy via various perturbations}
Based on the large-$J$ analysis, we find that the SO phase and the four decoupled phase are infinitely degenerate.
The presence of infinite degeneracies arises from two aspects. One is from the arbitrary spin moment in a SO bond.
Another is from the continuous rotation between the two decoupled lattices.
We will show that the massive degeneracy in the large-$J$ phases can be fully or partially lifted
due to the perturbation of various secondary interactions. The results are summarized in Table~\ref{Table:lift}.
We present the detailed analysis for each degenerate large-$J$ phase in the following.
%%%%%%%%%%%%%%%%%%%%%%%%%%%%%%%%
\begin{table}
\caption{\label{Table:lift}
Lift of the infinite degenerate ground states under various perturbed interactions.
}
\begin{ruledtabular}
\begin{tabular}{ccc}
Degenerate manifold    &  Perturbation & Resultant Phase \\ \hline
\multirow{4}{*}{Semi-ordered ($J_1^Q<0$)} &   $J^D_2\rightarrow 0^+$   & Diagonal FQ+AFM   \\
                              &   $J^D_2\rightarrow 0^-$   & Diagonal FQ+FM    \\
                              &   $J^Q_2\rightarrow 0^+$   & N\'{e}el AFQ   \\
                              &   $J^Q_2\rightarrow 0^-$   & Degenerate AFQ    \\ \hline
\multirow{2}{*}{Decoupled AFM ($J_2^D>0$)}&   $J^Q_1\rightarrow 0^+$   & CAFM  \\
                              &   $J^Q_1\rightarrow 0^-$   & OM    \\ \hline
\multirow{4}{*}{Decoupled FM ($J_2^D<0$)} &   $J^D_1\rightarrow 0^+$   & AFM   \\
                              &   $J^D_1\rightarrow 0^-$   & FM    \\
                              &   $J^Q_1\rightarrow 0^+$   & FM/AFM\\
                              &   $J^Q_1\rightarrow 0^-$   & OM    \\ \hline
\multirow{2}{*}{Decoupled FQ ($J_2^Q>0$)} &   $J^Q_1\rightarrow 0^+$   & FQ  \\
                              &   $J^Q_1\rightarrow 0^-$   & N\'{e}el AFQ    \\ \hline
\multirow{4}{*}{Decoupled SO ($J_2^Q<0$)} &   $J^D_1\rightarrow 0^+$   & Stripe FQ+AFM   \\
                              &   $J_1^D\rightarrow 0^-$   & Stripe FQ+FM    \\
                              &   $J_1^Q\rightarrow 0^+$   & Decoupled AFQ   \\
                              &   $J_1^Q\rightarrow 0^-$   & Degenerate SO    \\
\end{tabular}
\end{ruledtabular}
\end{table}
%%%%%%%%%%%%%%%%%%%%%%%%%%%%%%%%

(i) Lift of degeneracy in the SO phase.
Since the expectation value of a pair of spin operators can be written as
\begin{align}
 \langle\bs_i\cdot\bs_j\rangle=|\dd_i^\ast\cdot\dd_j|^2-|\dd_i\cdot\dd_j|^2,
\end{align}
we see that $\langle\bs_i\cdot\bs_j\rangle$ is always zero when one of the two sites features a purely quadrupolar state.
Thus including the NN dipolar coupling $J_1^D$ has no consequence on the ground state energy of the SO phase.

\begin{figure}
\centering\includegraphics[scale=0.38]{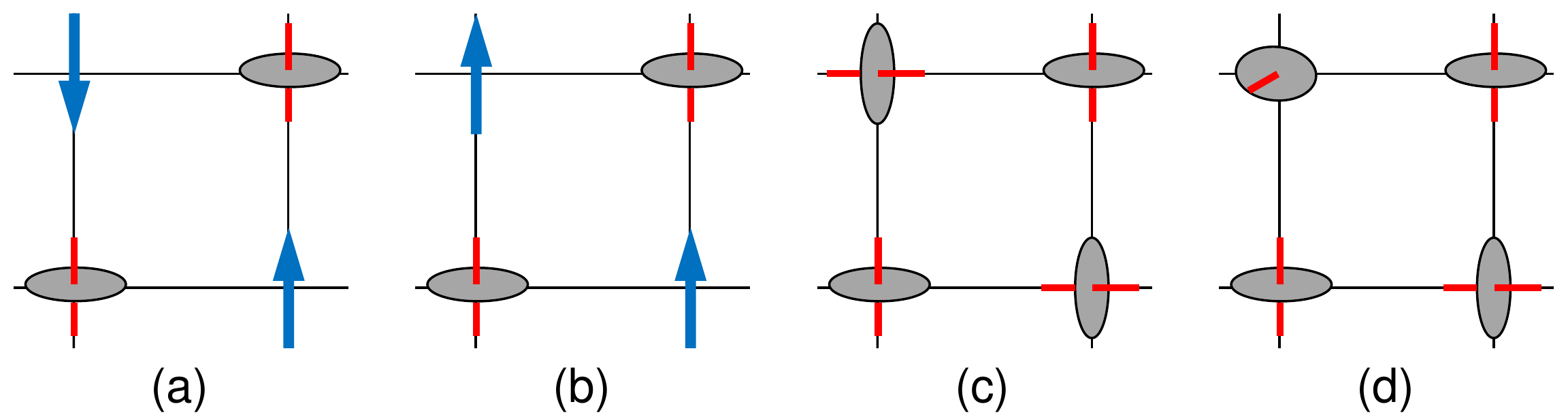}
\caption{(Color online)
Lift of the infinite degeneracies in the SO phase under various perturbed interactions leads to
(a) diagonal FQ+AFM for $J_2^D\rightarrow0^+$,
(b) diagonal FQ+FM for $J_2^D\rightarrow0^-$,
(c) N\'{e}el AFQ for $J_2^Q\rightarrow0^+$, and
(d) degenerate AFQ for $J_2^Q\rightarrow0^-$.
}
\label{fig:SOlift}
\end{figure}

However, we find that the degeneracy of the SO phase will be fully lifted in the presence of finite $J_2^D$ coupling ,
leading to two interpenetrating sublattices along the diagonal direction with one sublattice featuring FQ order
and the other featuring AFM ($J_2^D>0$) or FM ($J_2^D<0$) order, see Fig.~\ref{fig:SOlift} (a) and (b) for an
illustration on a plaquette. We call this phase diagonal FQ+FM for $J_2^D<0$ and diagonal FQ+AFM for $J_2^D>0$.
%%%%%%%%%%%%%%%%%%%%%%%%%%%%%%%%
%%%%%%%%%%%%%%%%%%%%%%%%%%%%%%%%

Likewise, the presence of finite $J_2^Q$ coupling will also lift the degeneracy of the SO phase.
For $J_2^Q>0$, the degeneracy is fully lifted,
leading to a two-sublattice N\'{e}el type AFQ order, see Fig.~\ref{fig:SOlift} (c).
However the massive degeneracy is partially lifted for $J_2^Q<0$ and we find that
a highly degenerate state with purely quadrupolar nature is stabilized.
This degenerate phase can be constructed by filling the square lattice with three purely
quadrupolar state (e.g. $\ket{x}$ , $\ket{y}$ and $\ket{z}$) by requiring all NN bonds featuring orthogonal state,
see Fig.~\ref{fig:SOlift} (d) for a sketch on a plaquette.
We call this phase degenerate AFQ.
In fact, the ground state configuration for the degenerate AFQ phase on the square lattice has been extensively studied
with only NN bilinear and biquadratic interactions~\cite{Papanicolaou1988367} and it is recently proposed that a peculiar
three-sublattice ordering is selected by quantum fluctuations~\cite{PhysRevLett.105.265301,PhysRevB.85.140403}.

\begin{figure}
\centering\includegraphics[scale=0.4]{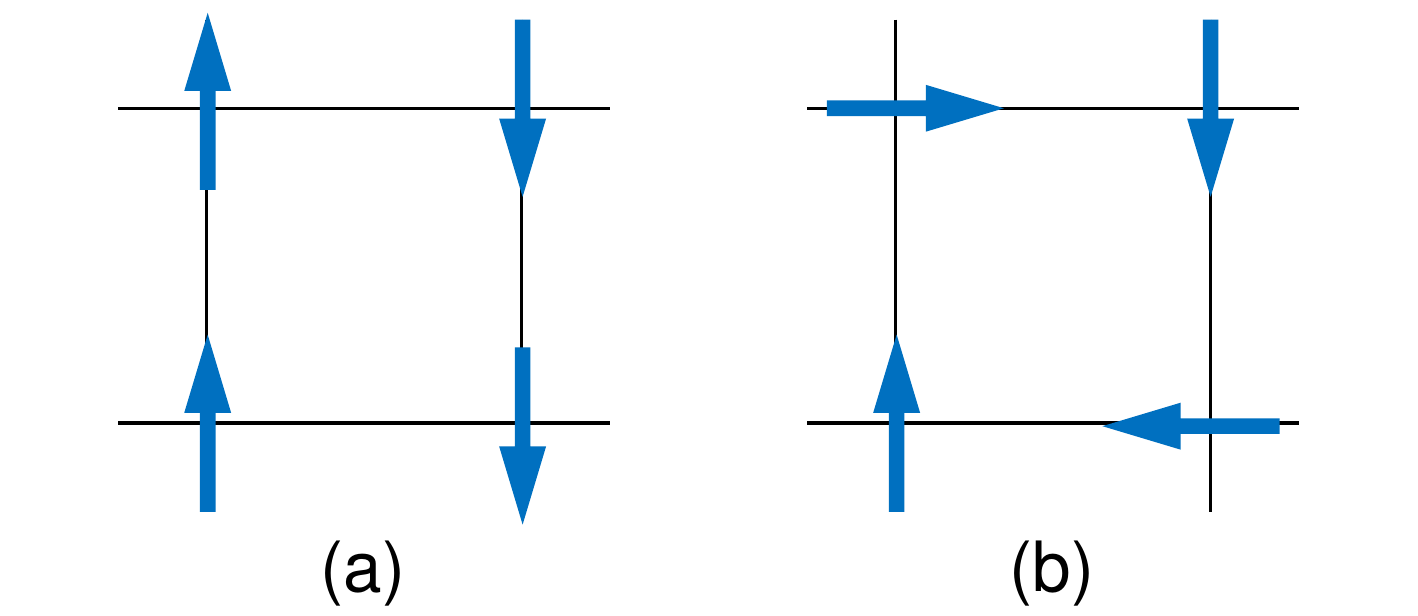}
\caption{(Color online)
Illustration of two magnetic phases exhibiting purely spin dipolar ordering.
(a) Collinear AFM (CAFM) phase with ordering wave vector $(\pi,0)$.
(b) Orthomagnetic (OM) phase without breaking the tertragonal lattice symmetry.
}
\label{fig:DAFMlift}
\end{figure}

(ii) Lift of degeneracy in the decoupled AFM phase.
It can be shown that only $J_1^Q$ perturbation has an impact on the ground state energy.
We find that the degeneracy due to the continuous rotation between the two decoupled lattices
will be fully lifted in the presence of a finite NN quadrupolar coupling,
leading to a collinear configuration (CAFM phase) with NN spin moments being parallel for $J^Q_1>0$ and an orthomagnetic (OM) phase with NN spin moments being perpendicular for $J^Q_1<0$.
The collinear phase has a twofold degeneracy with ordering wave vector
$(\pi,0)$ or $(0,\pi)$ while the OM phase preserves the tertragonal lattice symmetry~\cite{PhysRevB.91.024401}.
We depict these two phases in Fig.~\ref{fig:DAFMlift}(a) and (b), respectively.
%%%%%%%%%%%%%%%%%%%%%%%%%%%%%%%%
%%%%%%%%%%%%%%%%%%%%%%%%%%%%%%%%

(iii) Lift of degeneracy in the decoupled FM phase.
It can be shown that the decoupled FM phase is unstable with the perturbation of
all NN dipolar and quadrupolar couplings. The selection mechanism is strongly dependent on the sign of the interactions.
In particular, we find that the FM phase is selected by $J_1^D<0$, the AFM phase is selected by  $J_1^D>0$ and the OM phase is selected
by $J_1^Q<0$. However, for $J_1^Q>0$ both the FM and AFM phase are the variational ground states.

(iv) Lift of degeneracy in the decoupled FQ phase. We find that the degeneracy in the decoupled FQ phase is only lifted by
the presence NN quadrupolar couplings.
A uniform FQ order is selected by $J_1^Q>0$ while the N\'{e}el AFQ order is selected by $J_1^Q<0$.

(v) Lift of degeneracy in the decoupled SO phase.
The decoupled SO phase is unstable with the perturbation of all NN dipolar and quadrupolar couplings.

\begin{figure}
\centering\includegraphics[scale=0.38]{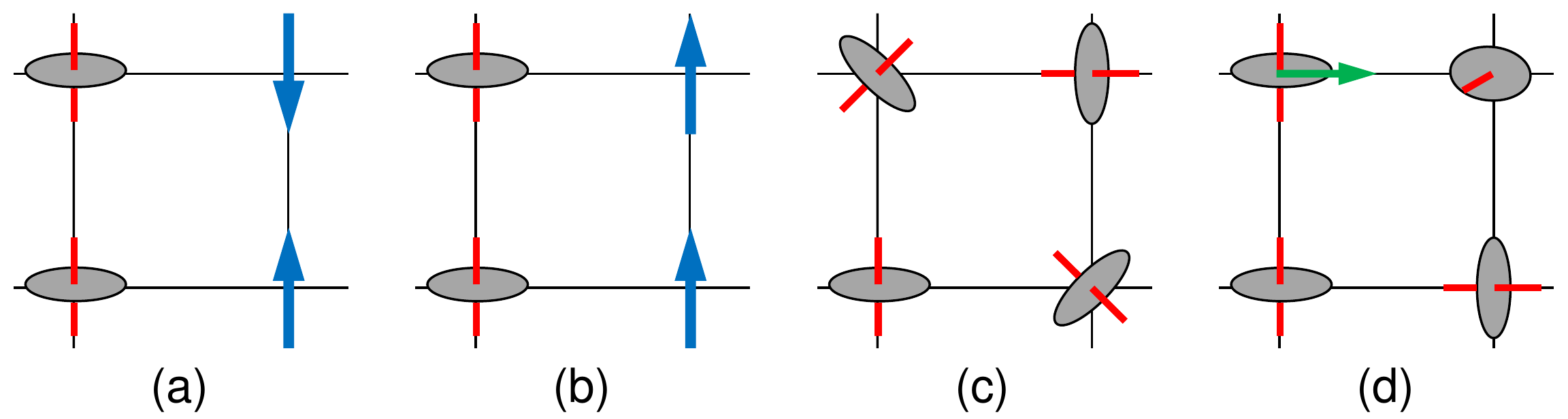}
\caption{(Color online)
Lift of the infinite degeneracies in the decoupled SO phase under various perturbed interactions leads to
(a) stripe FQ+AFM for $J_2^D\rightarrow0^+$,
(b) stripe FQ+FM for $J_2^D\rightarrow0^-$,
(c) decoupled AFQ for $J_2^Q\rightarrow0^+$,
(d) degenerate SO for $J_2^Q\rightarrow0^-$.
}
\label{fig:DSOlift}
\end{figure}

In the presence of perturbed NN dipolar interaction $J_1^D$,
one can obtain two different phases with mixed dipolar and quadrupolar characters.
In particular, one finds the ground state manifold will feature alternate dipolar and quadrupolar
alignments along the columns (rows) with collinear directors and moments.
Since the directors of the quadrupolar alignment should be parallel to their adjacent spin moments,
we denote this phase as stripe FQ+AFM for $J_1^D>0$ and stripe FQ+FM for $J_1^D<0$.
An illustration of the two phases on a plaquette is depicted in Fig.~\ref{fig:DSOlift}(a) and (b).
Notice that the two stripe phases on the square lattice are still infinitely degenerate
since every AFM or FM column (row) is decoupled. Thus the degeneracy is only partially lifted in this case.
%%%%%%%%%%%%%%%%%%%%%%%%%%%%%%%%
%%%%%%%%%%%%%%%%%%%%%%%%%%%%%%%%

In the presence of perturbed NN quadrupolar interactions, one still obtains two highly degenerate phases.
For $J_1^Q>0$, the variational solution supports a peculiar purely quadrupolar phase
which consists of two decoupled N\'{e}el AFQ lattices, see Fig.~\ref{fig:DSOlift}(c).
In this case, it is expected that a quantum order-by-disorder mechanism can break the continuous degeneracy
due to the arbitrary angle between the staggered directors of these two sublattices, leading to a FQ alignment
along one direction with parallel directors between two NN sites and an alternate AFQ configuration
along the the other direction with orthogonal directors between two NN sites.
We note that the effect of additional third NN interactions is important to stabilize a genuine
two-sublattice collinear AFQ (CAFQ) phase with ordering wave vector $(\pi,0)/(0,\pi)$.
The $(\pi,0)$ AFQ order is recently proposed to explain the curious magnetic properties and nematicity of FeSe,
see Ref.~\cite{PhysRevLett.115.116401}.
For $J_1^Q<0$, the degeneracy of one diagonal SO bond on a plaquette will be lifted,
leading to a purely quadrupolar configuration with orthogonal directors,
while the the degeneracy of the other diagonal SO bond is still preserved, see Fig.~\ref{fig:DSOlift}(d)
for an illustration. We denote this phase as degenerate SO.

\subsection{Phase diagram relevant to iron-based SCs}

Based on the above variational analysis, we find that the extended BBQ model (\ref{BBQmodel}) can
support various ground state manifolds including conventional spin dipolar orders (ferro- and antiferromagnet),
novel quadrupolar orders (spin nematic) and mixed dipolar-quadrupolar orders.
An exhaustive phase diagram can be mapped out by comparing the ground state energy of different phases.
Since we are interested in the emergence of possible new ground states relevant to iron-based SCs,
we restrict our discussion to $J_1^Q>0$ (or $K_1>0$) regime where the $(\pi,0)$ CAFM phase
for iron pnictides is stabilized in the presence of dominant antiferromagnetic $J_2$ interaction.
Henceforth, we will set $J_2=1$ as the energy unit in order to incorporate the NNN antiferromagnetic
superexchange processes in iron pnictides and chalcogenides.

\begin{figure*}
\centering\includegraphics[scale=0.63]{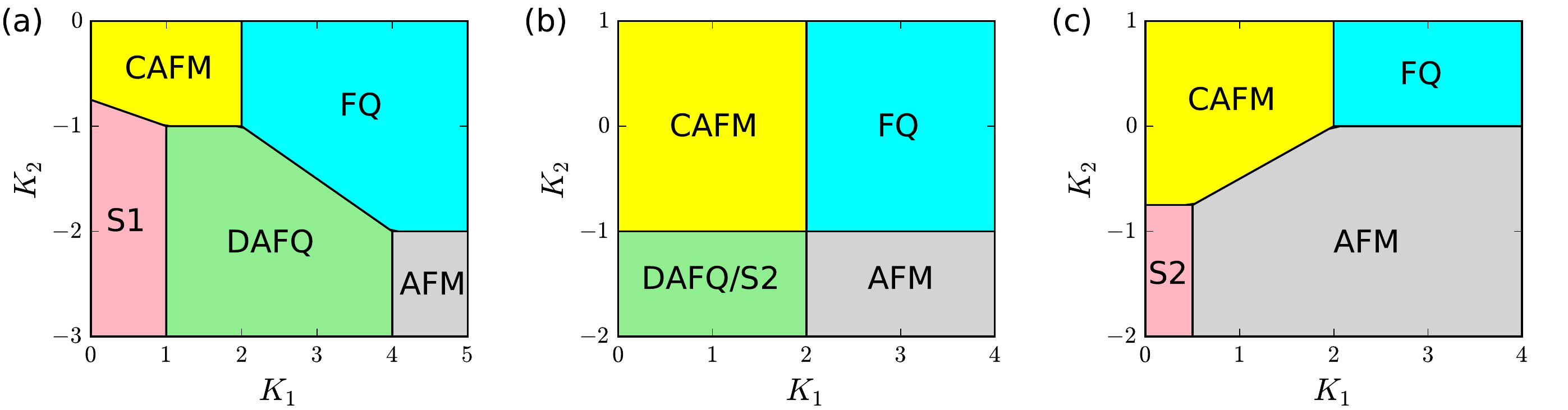}
\caption{(Color online)
The variational ground-state phase diagram for the spin-1 bilinear-biquadratic (BBQ) model (\ref{BBQmodel})
on the square lattice. We have set $J_2=1$ and (a) $J_1=-1$, (b) $J_1=0$, (c) $J_1=1$.
S1 and S2 represent stripe FQ+FM and stripe FQ+AFM phases, respectively.
}
\label{fig:Phasediagram}
\end{figure*}

A portion of the variational phase diagram under the influence of variable
$K_1$ and $K_2$ interactions is mapped out in Fig.~\ref{fig:Phasediagram} for several $J_1$ interactions.
%%%%%%%%%%%%%%%%%%%%%%%%%%%%%%%%
%%%%%%%%%%%%%%%%%%%%%%%%%%%%%%%%
We see that for strong NN biquadratic coupling $K_1$ the ground state consists of FQ or AFM order since
$K_1$ enhances both antiferromagnetic dipolar coupling as well as ferroquadrupolar coupling
among NN sites according to Eq.~(\ref{BQterm}). The CAFM phase occupies the upper left part of the
phase diagram since a large and positive $K_2$ interaction will enhance the NNN antiferromagnetic dipolar coupling
and the presence of moderate $K_1>0$ will select the collinear phase at the variational level.
The most interesting feature is in the lower left part of the phase diagram.
We find that for $J_1<0$ it is occupied by a mixed dipolar-quadrupolar phase (stripe FQ+FM)
and a purely quadrupolar phase (decoupled AFQ).
While for $J_1\geq0$ the stripe FQ+FM disappears from the present phase diagram.
Specifically, both DAFQ and another mixed phase (stripe FQ+AFM) are the ground state for $J_1=0$ and the latter
dominates the whole lower left region for $J_1>0$.
Thus the emergence of a nonuniform purely quadrupolar phase in the vicinity of the CAFM phase, including the proposed $(\pi,0)$ AFQ phase relevant to FeSe as selected by quantum fluctuations in the DAFQ phase,
is most likely to be realized for $J_1<0$.

%%%%%%%%%%%%%%%%%%%%%%%%%%%%%%%%%%%%%%%%%%%%%%%%%%%%%%%%%%%%%%%%
\section{SU(3) flavor-wave theory}\label{Sec.flavor wave}
%%%%%%%%%%%%%%%%%%%%%%%%%%%%%%%%%%%%%%%%%%%%%%%%%%%%%%%%%%%%%%%%
Since the quadrupolar operators are related to the generators of SU(3) Lie algebra, the spin dipolar
and quadrupolar order parameters fluctuate in the SU(3) space instead of the SU(2) space of
local spin rotations~\cite{Batista10.1080,RevModPhys.86.563}.
The SU(3) flavor-wave theory for spin-1 systems starts from introducing three-flavor Schwinger bosons (SBs)
$a_{\mu}^\dag$ (with $\mu=1,2,3$) which create the three local spin-1 basis
$a_{\mu}^\dag\ket{\emptyset}=\ket{\mu}$
and satisfy the local constraint~\cite{PhysRevB.88.184430,Muniz01082014,PhysRevB.91.174402}
\begin{align}\label{constraint}
\sum_\mu a_\mu^\dag a_\mu=1.
\end{align}
For the description of conventional spin dipolar phases, it is convenient to use the usual $S^z$ basis
$\ket{\boldsymbol{\mu}}=(\ket{\bar{1}},\ket{0},\ket{1})$, while the time-reversal invariant basis
$\ket{\boldsymbol{\mu}}=(\ket{x},\ket{y},\ket{z})$ is used for the study of quadrupolar phases.
In terms of the SBs, the local spin and quadrupolar operators have bilinear forms which are constructed by
the eight generators of SU(3) group in the fundamental representation~\cite{Auerbach2012,PhysRevB.84.054406}.
In the next we investigate the dynamic properties of the BBQ model in the $(\pi,0)$ CAFM phase
within the framework of SU(3) flavor-wave theory.

\subsection{General results for the $(\pi,0)$ CAFM phase}

\begin{figure}
\centering\includegraphics[scale=0.6]{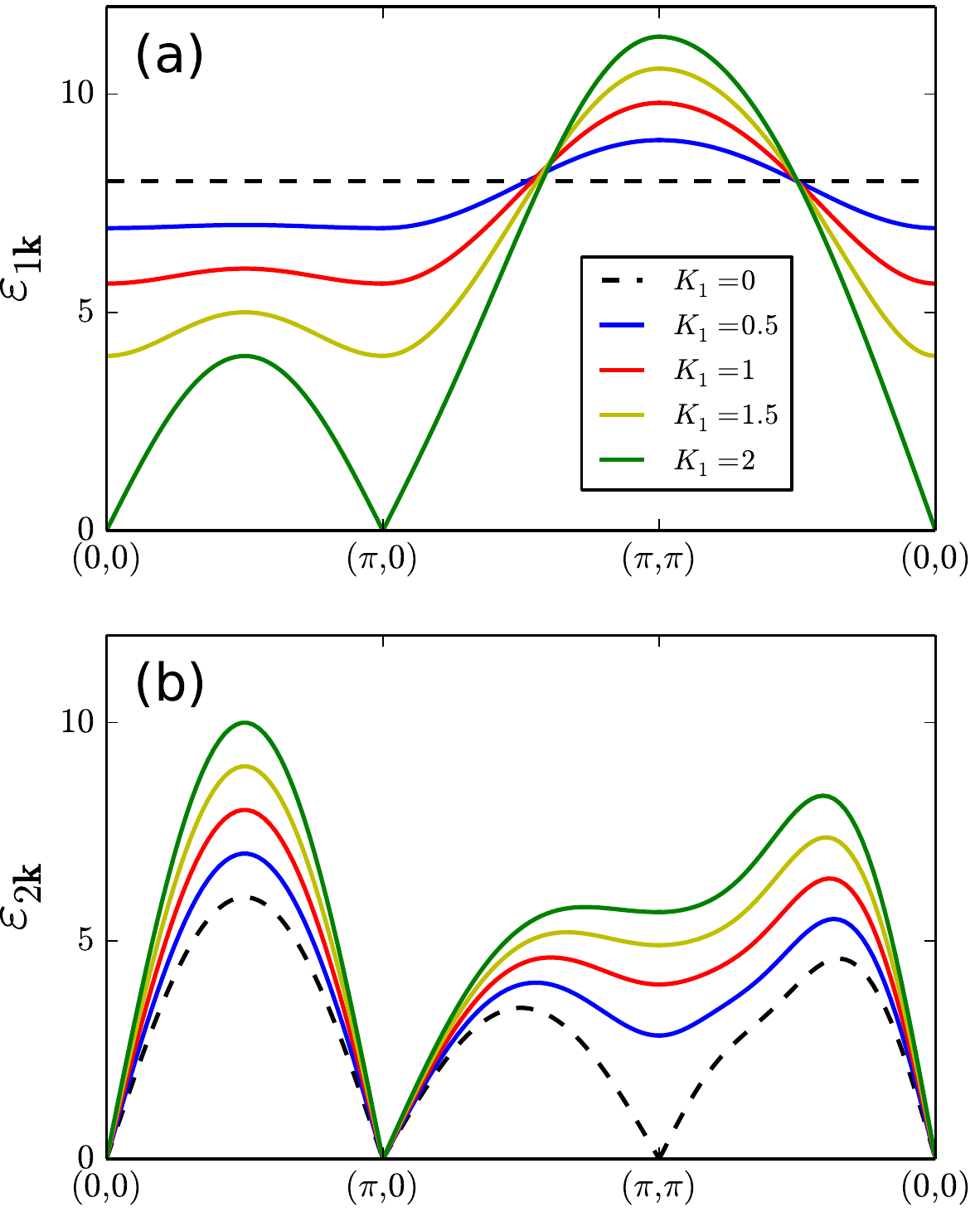}
\caption{(Color online)
Elementary excitation spectra of the flavor waves in the CAFM phase of the BBQ model (\ref{BBQmodel})
with $J_1=J_2=1$ and $K_2=0$.
(a) The high-energy (quarupolar) branch is gapped corresponding to the elementary $\Delta S=2$ excitations.
(b) The low-energy spectra are gapless magnon mode corresponding to the elementary $\Delta S=1$ excitations.
}
\label{fig:flavordispersion}
\end{figure}

The three-flavor SBs which create the three local spin-1 basis are introduced as
\begin{align}\label{cafmSBs}
  a^\dag_1\ket{\emptyset}=\ket{\bar{1}},\ a^\dag_2\ket{\emptyset}=\ket{\bar{0}},\ a^\dag_3\ket{\emptyset}=\ket{1},
\end{align}
and satisfy the local constraint (\ref{constraint}).
In terms of the three-flavor SBs, the local spin and quadrupolar operators are constructed by
the fundamental representation of SU(3) Lie algebra which have the following bilinear
forms~\cite{Batista10.1080,RevModPhys.86.563}
\begin{align}\label{su3operators}
  \left(
\begin{array}{c}
  S^x \\
  S^y \\
  S^z \\
  Q^{x^2-y^2} \\
  Q^{3z^2-r^2} \\
  Q^{xy} \\
  Q^{yz} \\
  Q^{zx}
\end{array}\right)=
\left(
\begin{array}{c}
  \frac{1}{\sqrt{2}}(a^\dag_1 a_2+a_2^\dag a_3+\mathrm{H.c.})\\
  \frac{i}{\sqrt{2}}(a^\dag_1 a_2+a_2^\dag a_3-\mathrm{H.c.}) \\
  a_3^\dag a_3-a_1^\dag a_1 \\
  a_1^\dag a_3+a_3^\dag a_1 \\
  \frac{1}{\sqrt{3}}(1-3a_2^\dag a_2) \\
  i(a_1^\dag a_3-a_3^\dag a_1) \\
  \frac{i}{\sqrt{2}}(a_2^\dag a_3+a_2^\dag a_1-\mathrm{H.c.}) \\
  \frac{1}{\sqrt{2}}(a_2^\dag a_3-a_2^\dag a_1+\mathrm{H.c.})
\end{array}\right).
\end{align}
In order to study the two-sublattice CAFM phase,
it is convenient to perform a local rotation in the spin space
\begin{align}\label{rotation}
  S_i^{x_0}=e^{i\boldsymbol{\kappa}\cdot\rr_i}S_i^{x},\ S_i^{y_0}=S_i^y,
\ S_i^{z_0}=e^{i\boldsymbol{\kappa}\cdot\rr_i}S_i^{z},
\end{align}
where $\boldsymbol{\kappa}=(\pi,0)$ is the corresponding ordering wave vector.
In the local frame, the following mean-field ground state is stabilized
\begin{align}
 \ket{\Psi}=\prod_{i=1}^Na_{3i}^\dag\ket{\emptyset}.
\end{align}
The leading quantum correction above the mean-field ground state is described by the SU(3) flavor-wave theory
which is implemented via the condensation of the boson $a_{3}$ under the local constraint (\ref{constraint})
\begin{align}\label{hpexpansion}
  a_{3}^\dag, a_{3}\rightarrow\sqrt{1-a_1^\dag a_1-a_2^\dag a_2}.
\end{align}
The SB operators $(a_1,a_2)$ thus play the role of the Holstein-Primakoff bosons
in the SU(2) spin-wave theory which describe the fluctuations around the variational ground state.
After performing the Fourier transformation, we obtain the quadratic SU(3) flavor-wave Hamiltonian
\begin{align}
 \mathcal{H}=\sum_{\kk,\nu=1,2}A_{\nu\kk}a_{\nu\kk}^\dag a_{\nu\kk}
 +\frac{B_{\nu\kk}}{2}(a_{\nu\kk}a_{\nu-\kk}+\mathrm{H.c.}),
\end{align}
where
\begin{align}
  A_{1\kk}=&8J_2-2K_1\cos k_y+4K_2,\\
  B_{1\kk}=&2K_1\cos k_x-4K_2\cos k_x\cos k_y,\\
  A_{2\kk}=&2J_1\cos k_y+4J_2+2K_1+4K_2,\\
  B_{2\kk}=&-2(J_1+K_1)\cos k_x-4(J_2+K_2)\cos k_x\cos k_y.
\end{align}
The resulting Hamiltonian can be diagonalized via a Bogoliubov transformation
\begin{align}
 a_{\nu\kk}=u_{\nu\kk} b_{\nu\kk}+v_{\nu\kk} b_{\nu-\kk}^\dag,
\end{align}
with
\begin{align}
u_{\nu\kk}^2,v_{\nu\kk}^2=\frac{A_{\nu\kk}\pm\varepsilon_{\nu\kk}}{2\varepsilon_{\nu\kk}},
u_{\nu\kk}v_{\nu\kk}=-\frac{B_{\nu\kk}}{2\varepsilon_{\nu\kk}}.
\end{align}
The elementary excitation spectrum consists of two branches
\begin{align}\label{CAFdisp}
 \varepsilon_{\nu\kk}=\sqrt{A_{\nu\kk}^2-B_{\nu\kk}^2}.
\end{align}
According to (\ref{cafmSBs}), we find that the operator $a^\dag_1$ creates an excitation with $\Delta S^z=2$,
the corresponding branch $\varepsilon_{1\kk}$ has a quadrupolar (nematic) character,
which is the bound state of two magnons and always gapped in the magnetic dipolar phase.
While $a^\dag_2$ creates an excitation with $\Delta S^z=1$, the corresponding branch
$\varepsilon_{2\kk}$ is the conventional magnon mode (Goldstone mode) with gapless excitations.
The quasiparticle dispersion for several representative points in the phase diagram of the CAFM phase
is shown in Fig.~\ref{fig:flavordispersion}.
%%%%%%%%%%%%%%%%%%%%%%%%%%%%%%%%
%%%%%%%%%%%%%%%%%%%%%%%%%%%%%%%%
Generally, the high-energy quadrupolar branch forms a perfectly flat band with a finite gap
in the absence of biquadratic couplings, see Fig.~\ref{fig:flavordispersion}(a).
While we see that when the system approaches a nematic critical point, as shown in Fig.~\ref{fig:flavordispersion}(a),
the quadrupolar branch eventually becomes gapless excitations, signaling the onset of a quadrupolar order.
It is already clear from Eq.~(\ref{CAFdisp}) that the low-energy magnon spectrum has
zeros at $\kk=(0,0)$ and $\kk=(\pi,0)$ in the two-dimensional Brillouin zone, see Fig.~\ref{fig:flavordispersion}(b).
It is also shown in Fig.~\ref{fig:flavordispersion}(b) that the presence of finite NN biquadratic coupling $K_1$ opens the spin-wave gap at momentum $(\pi,\pi)$.

\subsection{Spin-wave dispersion of iron pnictides}

We now discuss the applicability of the extended BBQ model for iron pnictides and determine
which regime in the parameter space is the most relevant to the experimentally observed spin-wave dispersion.
Experimentally, the magnon dispersion is best known for the 122 compounds CaFe$_2$As$_2$~\cite{NatPhys.5.555}
and BaFe$_2$As$_2$~\cite{PhysRevB.84.054544}.
Hence we fit the dispersion with the SU(3) flavor-wave theory of
the BBQ model (\ref{BBQmodel}) to the measured spin-wave dispersion for iron pnictides.
Then we are able to compare the exchange values with those obtained by the conventional SU(2) spin-wave approaches
(see Appendix~\ref{App:SU(2)SWT} for a summary of the spin-wave theory).

\begin{figure}
\centering\includegraphics[scale=0.43]{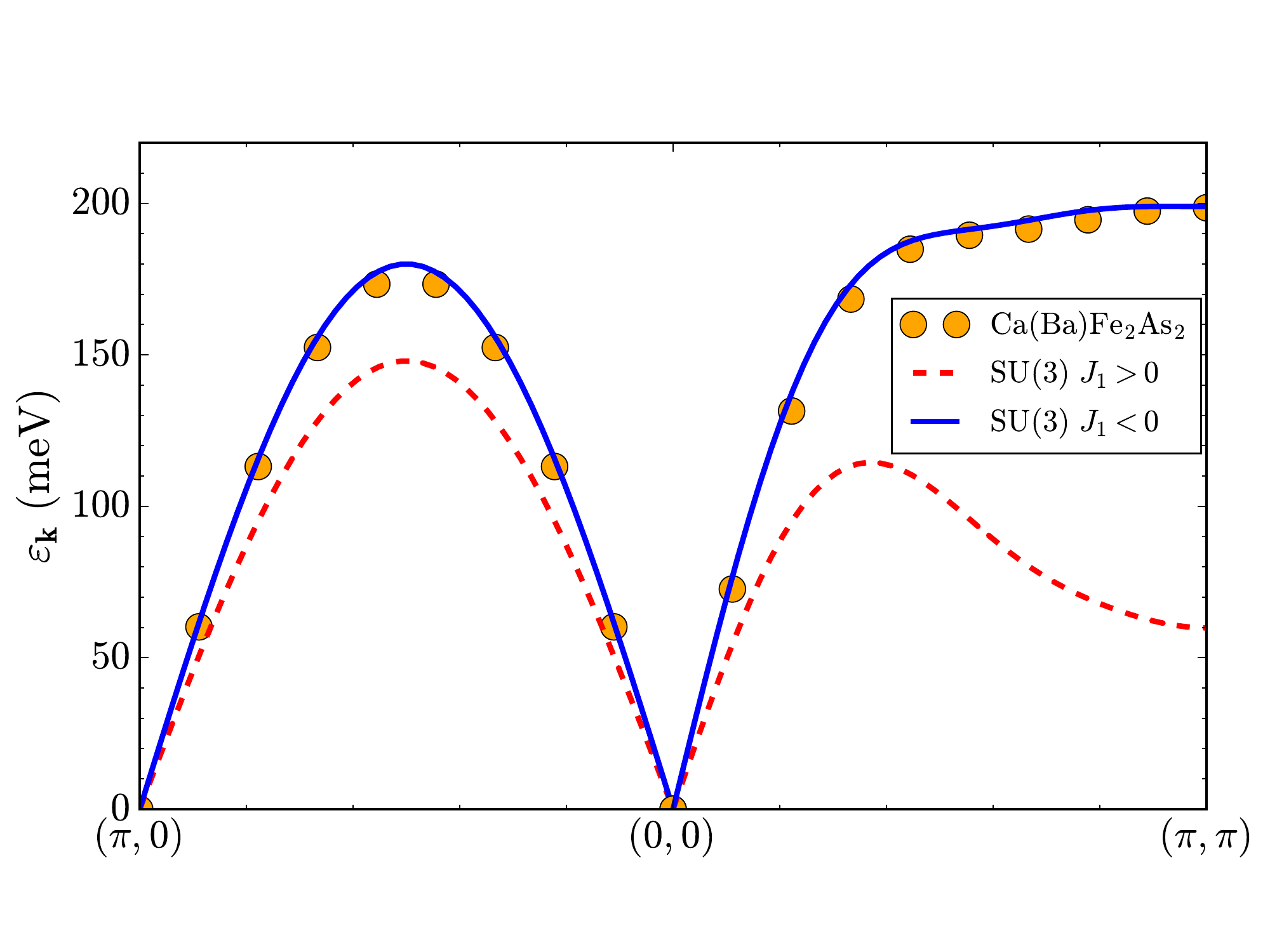}
\caption{(Color online)
Spin-wave spectrum for iron pnictides Ca(Ba)Fe$_2$As$_2$.
The orange circles represent the experimental dispersion fitted by a $J_x-J_y-J_2$ model
which is equivalent to the $J_1-J_2-K_1$ model in the linear SU(2) spin-wave theory
with the accepted parameters $J_1=22$ meV, $J_2=19$ meV and $K_1=14$ meV~\cite{NatPhys.5.555,PhysRevB.84.054544}.
The red dashed curve represents the dispersion plotted by the same set of parameters but in the framework of
SU(3) flavor-wave theory.
The best fitting of the experimental dispersion with SU(3) flavor-wave theory as shown by the blue curve
is achieved with $J_1=-5$ meV, $J_2=50$ meV, $K_1=45$ meV
and $K_2=-25$ meV.
}
\label{fig:DispersionFit}
\end{figure}

To address this issue we would like to make a brief review of the previous efforts.
The experimentally observed INS spectrum of CAFM iron pnictides exhibits a striking feature, namely,
the spin-wave energy approximately forms a maximum
at $(\pi,\pi)$ point~\cite{NatPhys.5.555,PhysRevB.84.054544,PhysRevB.84.155108}.
This distinct feature can not explained by a simple Heisenberg $J_1-J_2$ model which predicts a minimum
at $(\pi,\pi)$ point~\cite{PhysRevB.78.052507,PhysRevB.79.092416,PhysRevB.83.144528}.
The initial fittings were performed by a phenomenological $J_x-J_y-J_2$ model
with antiferromagnetic $J_x$ and nearly ferromagnetic $J_y$,
which do not allow for a reconciliation even when the small orthorhombic lattice distortion is taken into account.
Therefore, it is of great importance to formulate a minimum spin model which can capture both
the correct spin-wave spectrum and preserve the tetragonal lattice symmetry.
Such efforts were subsequently made by the inclusion of a biquadratic coupling between NN sites
which is just the celebrated $J_1-J_2-K_1$ model.
The $J_1-J_2-K_1$ model studied in Refs.~\cite{NatPhys.7.485,PhysRevB.86.085148,PhysRevB.85.144403} is based on
a mean-field decoupling of the biquadratic term which is shown to be identical to the $J_x-J_y-J_2$ model,
see Appendix~\ref{App:SU(2)SWT} for a exact mapping between these two models under a Hubbard-Stratonovich
transformation. Thus such fitting gives the traditionally accepted magnetic exchange interactions with
$J_1=22$ meV, $J_2=19$ meV and $K_1=14$ meV for CaFe$_2$As$_2$, see the orange circles in Fig.~\ref{fig:DispersionFit}.
However, the limitations of the mean-field decoupling are realized subsequently by authors
in Refs.~\cite{PhysRevB.84.064505,PhysRevB.92.165102} where they carry out a nonlinear spin-wave
calculation and find that the experimental spectra can not be well captured even with fairly large biquadratic coupling.

%%%%%%%%%%%%%%%%%%%%%%%%%%%%%%%%
%%%%%%%%%%%%%%%%%%%%%%%%%%%%%%%%

We want to point out that all these fittings carried out in the previous works are based on the assumption
of $J_1\approx J_2$~\cite{NatPhys.7.485,PhysRevB.86.085148,PhysRevB.85.144403}.
However, a dominant antiferromagnetic $J_2$ coupling can be accounted for by taking into the arsenic bridging superexchange process~\cite{PhysRevLett.101.076401} and even a ferromagnetic $J_1$ coupling is proposed in some iron chalcogenides
(see Table I in Ref.~\cite{PhysRevB.85.144403} for a summary of the exchange constants).
Furthermore, the dramatic reduction or even the sign change of $J_1/J_2$ signals the importance of the $p$ orbitals of As or Te/Se on the influence of magnetism in iron-based SCs.
In the present work, we will extend the fitting by including an adjustable $J_1$ interaction ranging from ferromagnetic to antiferromagnetic in the framework of SU(3) flavor-wave theory.
We find that for $J_1\approx J_2$ scenario, the existence of local minimum at $(\pi,\pi)$ is robust against
both the NN and NNN biquadratic interactions which is in good agreement with the prediction by nonlinear SU(2) spin-wave theory ~\cite{PhysRevB.92.165102}.
The best fitting of the experimental magnon dispersion for Ca(Ba)Fe$_2$As$_2$ under $J_1\ll J_2$ scenario
in the framework of SU(3) flavor-wave theory is achieved for $J_1=-0.1J_2$, $K_1=0.9J_2$ and $K_2=-0.5J_2$
with the energy scale $J_2=50$ meV, see the blue curve Fig.~\ref{fig:DispersionFit}.
We also show that the previously accepted exchange constants that were obtained by conventional SU(2) spin-wave theory
fail to reproduce the experimental dispersion in the new framework of SU(3) flavor-wave theory,
see the red dashed curve in Fig.~\ref{fig:DispersionFit},
since the spin quadrupolar nature is neglected in conventional SU(2) spin-wave theory.

%%%%%%%%%%%%%%%%%%%%%%%%%%%%%%%%%%%%%%%%%%%%%%%%%%%%%%%%%%%%%%%%
\section{Dynamic correlation functions}\label{Sec.dynamic correlations}
%%%%%%%%%%%%%%%%%%%%%%%%%%%%%%%%%%%%%%%%%%%%%%%%%%%%%%%%%%%%%%%%

\subsection{Spin dynamic structure factors}

Neutron scattering cross section is directly related to the diagonal components
of the spin dipolar dynamical structure factor (DSF), or the dynamical spin-spin  correlation function
\begin{align}
  \mathcal{S}_D^{\alpha_0\beta_0}(\qq,\omega)=\int_{-\infty}^\infty\frac{\mathrm{d}t}{2\pi}e^{i\omega t}
\langle S^{\alpha_0}_\qq(t)S^{\beta_0}_{-\qq}(0)\rangle,
\end{align}
where $\alpha_0$ and $\beta_0$ refer to spin dipolar components in the laboratory frame $\{x_0,y_0,z_0\}$.
Using (\ref{rotation}), we obtain the spin dipolar DSF in the rotating frame
\begin{align}\label{rotatingDSF}
  \mathcal{S}_D^{\mathrm{tot}}(\qq,\omega)=\mathcal{S}^{xx}_{\qq-\boldsymbol{\kappa},\omega}+
\mathcal{S}^{yy}_{\qq,\omega}+\mathcal{S}^{zz}_{\qq-\boldsymbol{\kappa},\omega},
\end{align}
In Eq.~(\ref{rotatingDSF}) one can readily identify the conventional
transverse and longitudinal components of the DSF
which are respectively given by
\begin{align}\label{TandLDSF}
\mathcal{S}^{T}(\qq,\omega)=\mathcal{S}^{xx}_{\qq-\boldsymbol{\kappa},\omega}+\mathcal{S}^{yy}_{\qq,\omega},\
\mathcal{S}^{L}(\qq,\omega)=\mathcal{S}^{zz}_{\qq-\boldsymbol{\kappa},\omega}.
\end{align}
In order to determine the leading contributions of the one- and two-particle excitations
to the total DSF in (\ref{rotatingDSF}), the spin operators are expanded to the quadratic terms
according to (\ref{su3operators}) with the help of (\ref{hpexpansion}), leading to
\begin{align}\label{sexpand}
  \left(
\begin{array}{c}
  S^x \\
  S^y \\
  S^z
\end{array}\right)\approx
\left(
\begin{array}{c}
  \frac{1}{\sqrt{2}}(a^\dag_2+a_1^\dag a_2+\mathrm{H.c.})\\
  \frac{i}{\sqrt{2}}(a^\dag_2+a_1^\dag a_2-\mathrm{H.c.}) \\
  1-2a_1^\dag a_1-a_2^\dag a_2
\end{array}\right).
\end{align}

\begin{figure}
\centering\includegraphics[scale=0.45]{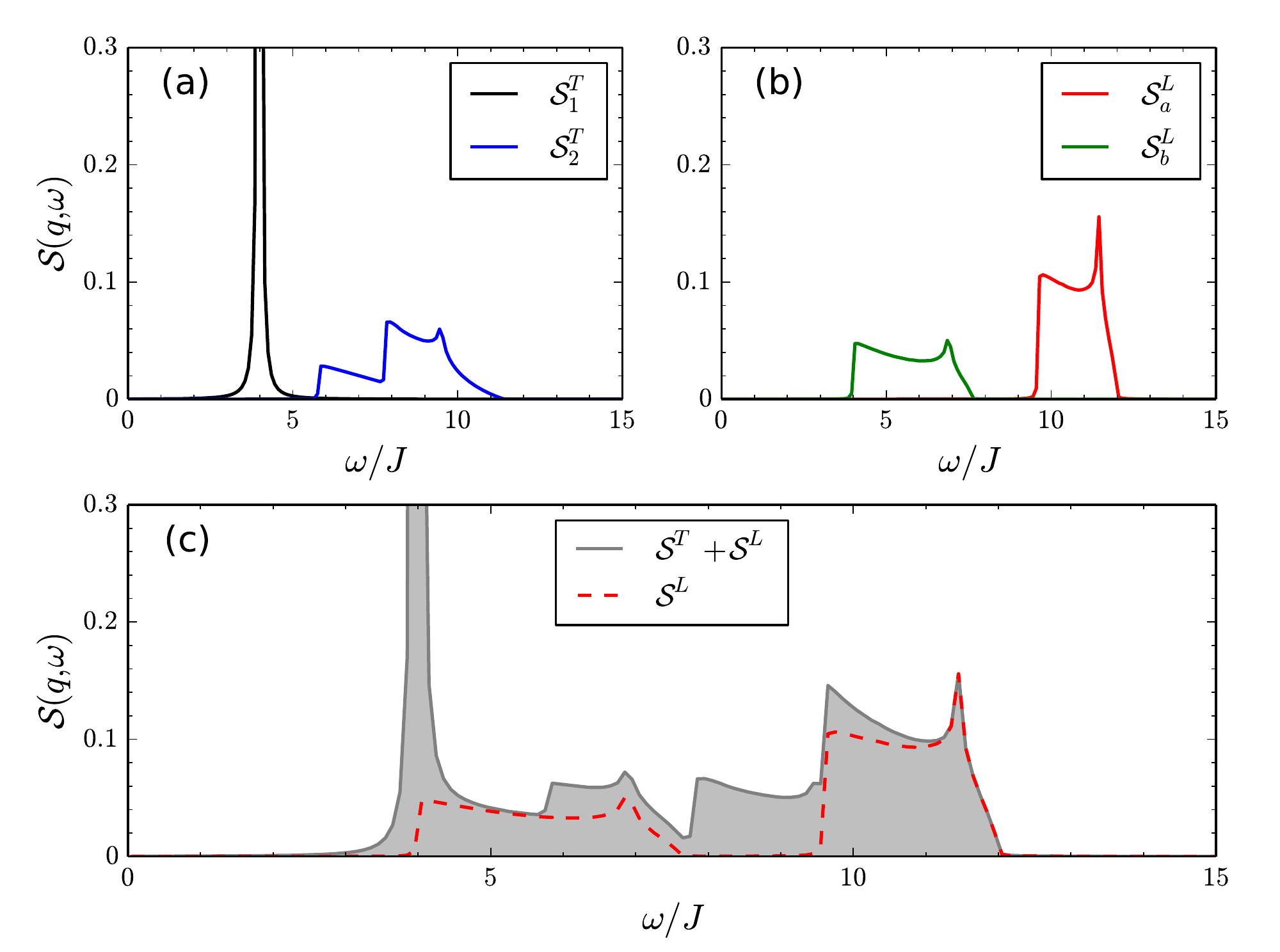}
\caption{(Color online)
Coherent and incoherent contributions to the spin dipolar dynamic structure factor (DSF) at momentum point $(\pi,\pi)$.
(a) The transverse DSF consists of both a coherent single-particle spectrum ($\mathcal{S}^{T}_1$)
and a two-particle continuum ($\mathcal{S}^{T}_2$).
(b) The longitudinal DSF consists of two incoherent two-particle continuum with contributions from two high-energy
quadrupolar modes ($\mathcal{S}^{L}_a$) and two low-energy magnon modes ($\mathcal{S}^{L}_b$).
(c) The total DSF ($\mathcal{S}^{T}+\mathcal{S}^{L}$) is shown by shaded areas,
while the total longitudinal component is plotted by red dashes lines.
}
\label{fig:DSFpipi}
\end{figure}

According to Eq.~(\ref{sexpand}) the transverse DSF can be expressed as
\begin{align}\label{TS-DSF}
\mathcal{S}^{T}(\qq,\omega)=\mathcal{S}^{T}_1(\qq,\omega)+\mathcal{S}^{T}_2(\qq,\omega),
\end{align}
with
\begin{align}
\mathcal{S}^{T}_1(\qq,\omega)&=(u_{2\qq}-v_{2\qq})^2\delta(\omega-\varepsilon_{2\qq}),\\
\mathcal{S}^{T}_2(\qq,\omega)&=\sum_{\kk+\kk^\prime=\qq}(u_{1\kk}v_{2\kk^\prime}-v_{1\kk}u_{2\kk^\prime})^2
\delta(\omega-\varepsilon_{1\kk}-\varepsilon_{2\kk^\prime}).
\end{align}
Apart from the coherent part $\mathcal{S}^{T}_1$ which is contributed by the single-magnon excitation,
one can clearly see that the transverse DSF also consists of a two-particle continuum.
The incoherent spectra $\mathcal{S}^{T}_2$ which can not be identified by SU(2) spin-wave theory
represents the simultaneous excitation of a magnon mode $\varepsilon_{2}$ plus a quadrupolar mode $\varepsilon_{1}$ .
In Fig.~\ref{fig:DSFpipi}(a) we present the transverse DSF at momentum point $(\pi,\pi)$ for CaFe$_2$As$_2$.
The coherent spectra show a $\delta$ function peak exactly at the spin-wave energy of $\omega\approx4J_2$,
while the high-energy incoherent continuum ranges from about $6J_2$ to $11J_2$.
Similarly, we find that the inelastic part of longitudinal DSF consists of two incoherent excitations
\begin{align}
\mathcal{S}^{L}(\qq,\omega)=\mathcal{S}^{L}_a(\qq,\omega)+\mathcal{S}^{L}_b(\qq,\omega),
\end{align}
with
\begin{align}
\mathcal{S}^{L}_a(\qq,\omega)&=\sum_{\kk+\kk^\prime=\qq}2(u_{1\kk}v_{1\kk^\prime}-v_{1\kk}u_{1\kk^\prime})^2
\delta(\omega-\varepsilon_{1\kk}-\varepsilon_{1\kk^\prime}),\\
\mathcal{S}^{L}_b(\qq,\omega)&=\sum_{\kk+\kk^\prime=\qq}\frac{1}{2}(u_{2\kk}v_{2\kk^\prime}-v_{2\kk}u_{2\kk^\prime})^2
\delta(\omega-\varepsilon_{2\kk}-\varepsilon_{2\kk^\prime}).
\end{align}
Thus the longitudinal DSF describes the excitations of two quadrupolar modes ($\mathcal{S}^{L}_a$)
and two magnon modes ($\mathcal{S}^{L}_b$) while the former can not be identified by SU(2) spin-wave theory either.
In Fig.~\ref{fig:DSFpipi}(b) we display the inelastic part of longitudinal DSF
at momentum point $(\pi,\pi)$  for CaFe$_2$As$_2$. It is interesting to point out that the two continua are
separated by a finite gap with $\mathcal{S}^{L}_{a(b)}$ contributing to high(low)-energy spectra.

%%%%%%%%%%%%%%%%%%%%%%%%%%%%%%%%
%%%%%%%%%%%%%%%%%%%%%%%%%%%%%%%%

The total DSF at momentum point $(\pi,\pi)$  for CaFe$_2$As$_2$ is also shown in Fig.~\ref{fig:DSFpipi}(c).
Though the two longitudinal continua $\mathcal{S}^{L}_{a}$ and $\mathcal{S}^{L}_{b}$ have no overlap,
the transverse continuum $\mathcal{S}^{T}_2$ bridges the gap and the three different continua give rise to
a robust high-energy sideband which may be verified experimentally.

\subsection{Quadrupolar dynamic structure factors}

Moreover, one will see that the correlation functions of spin quadrupolar operators will also come into play
even in the conventional magnetic dipolar phases.
To access the fingerprint of quadrupolar correlations, we consider the spin quadrupolar DSF
\begin{align}
  \mathcal{S}_Q(\qq,\omega)=\int_{-\infty}^\infty\frac{\mathrm{d}t}{2\pi}e^{i\omega t}
\langle \mathrm{Tr}[\bq_\qq(t)\bq_{-\qq}(0)]\rangle,
\end{align}
where we have kept the experimental details unspecified and calculate the quadrupolar DSF in the diagonalized representation.
Though the quadrupolar DSF can not be directly seen in conventional neutron probes
due to its nature of $\Delta S=2$ excitations,
it may be detected by optical measurements~\cite{PhysRevB.84.184424} under certain conditions and
it is recently proposed that the momentum resolved quadrupolar DSF can be experimentally discernible in
resonant inelastic x-ray scattering spectroscopy~\cite{arXiv:1506.04752}.

%%%%%%%%%%%%%%%%%%%%%%%%%%%%%%%%
\begin{figure}
\centering\includegraphics[scale=0.4]{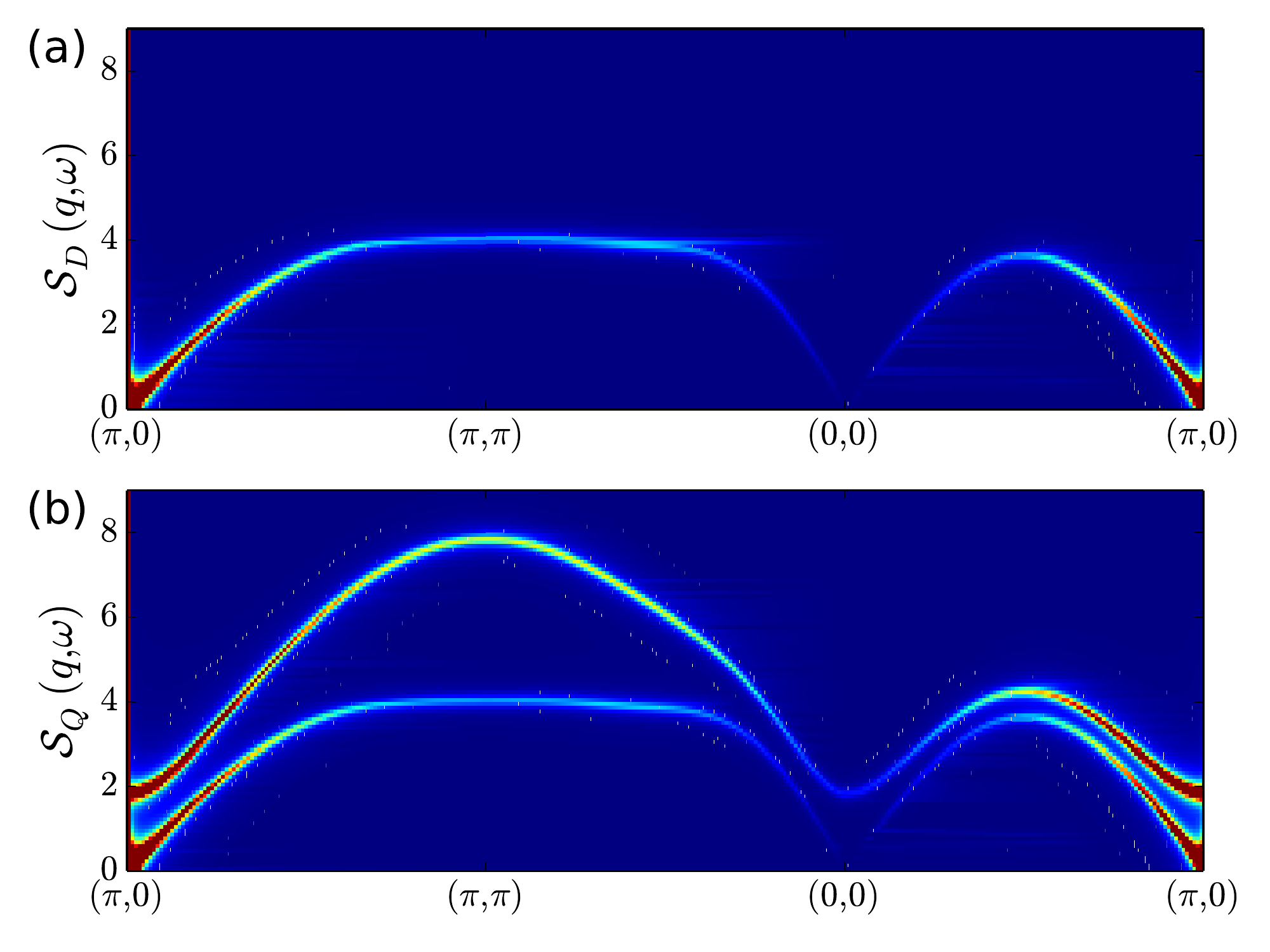}
\caption{(Color online)
The coherent excitations revealed by the spin dipolar (a)  and quadrupolar  (b)
dynamic structure factors in the CAFM phase of the BBQ model (\ref{BBQmodel})
with parameters fitted to Ca(Ba)Fe$_2$As$_2$.
}
\label{fig:DSFsq}
\end{figure}
%%%%%%%%%%%%%%%%%%%%%%%%%%%%%%%%

In order to determine the leading contributions of the coherent and incoherent excitations,
the quadrupolar operators can be also expanded to the quadratic terms
according to (\ref{su3operators}) with the help of (\ref{hpexpansion}), leading to
\begin{align}
  \left(
\begin{array}{c}
  Q^{x^2-y^2} \\
  Q^{3z^2-r^2} \\
  Q^{xy} \\
  Q^{yz} \\
  Q^{zx}
\end{array}\right)\approx
\left(
\begin{array}{c}
  a_1^\dag+a_1 \\
  \frac{1}{\sqrt{3}}(1-3a_2^\dag a_2) \\
  i(a_1^\dag-a_1) \\
  \frac{1}{\sqrt{2}}(a^\dag_2-a_1^\dag a_2-\mathrm{H.c.})\\
  \frac{i}{\sqrt{2}}(a^\dag_2-a_1^\dag a_2+\mathrm{H.c.})
\end{array}\right).
\end{align}
It is clear to see that the correlation function of $Q^{3z^2-r^2}$ can give rise to elastic scattering
cross section in experimental probes.
The presence of magnetic Bragg peak in quadrupolar DSF in the conventional magnets does not
signal any quadrupolar order parameter but exhibits the intrinsic properties of spin-1 systems.
The comparison of the spin dipolar and quadrupolar DSF of iron pnictides
is shown in Fig.~\ref{fig:DSFsq} where we have only calculated the dominated coherent spectra.
The low-energy magnon mode is revealed in the spin DSF with strong intensity near the antiferromagnetic
wave vector $(\pi,0)$ and gapless excitations at $(0,0)$, see Fig.~\ref{fig:DSFsq}(a).
However, both the low-energy magnon mode and high-energy quadrupolar mode are simultaneously revealed in the
quadrupolar DSF, see Fig.~\ref{fig:DSFsq}(b).
Note that in the quadrupolar DSF, the quadrupolar branch has a gap of about $2J_2$ which is directly accessible
by experimental measurements.

%%%%%%%%%%%%%%%%%%%%%%%%%%%%%%%%
\begin{figure}
\centering\includegraphics[scale=0.4]{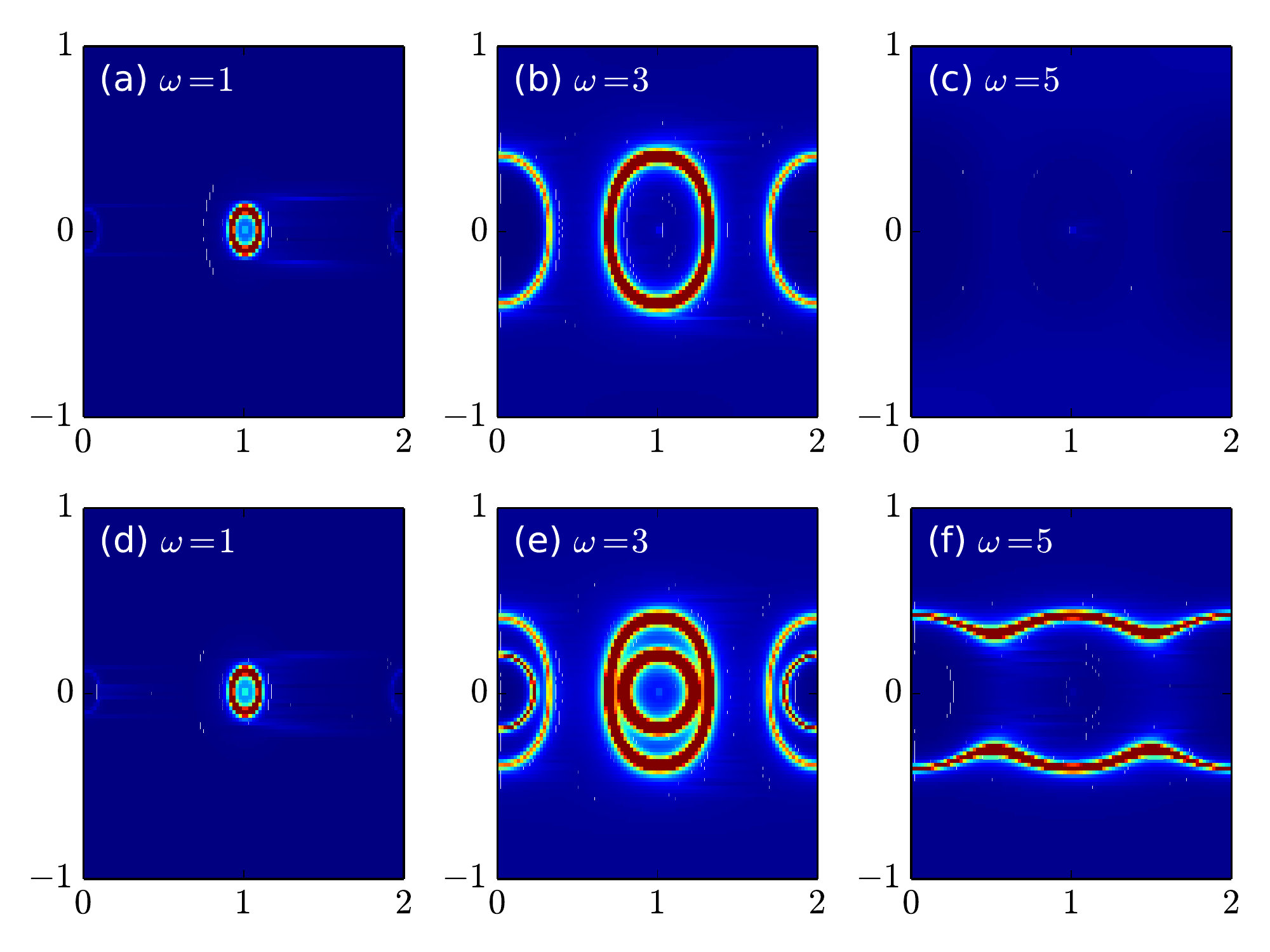}
\caption{(Color online)
Constant energy cuts of the coherent spin dipolar (a-c) and quadrupolar  (d-f)
dynamic structure factors for Ca(Ba)Fe$_2$As$_2$.
The cutting energies showing in the figures are in units of $J_2$ and
the reciprocal lattice vectors are given in units $\pi$/lattice constant.
}
\label{fig:DSFscan}
\end{figure}
%%%%%%%%%%%%%%%%%%%%%%%%%%%%%%%%

To better understand the spectral weight of DSFs in momentum space, we also present the
constant energy cuts of $\mathcal{S}_D(\qq,\omega)$ and $\mathcal{S}_Q(\qq,\omega)$ in Fig.~\ref{fig:DSFscan}.
At low energies, as displayed in Figs.~\ref{fig:DSFscan}(a) and (d),
the spin dipolar and quadrupolar DSF show similar structure with elliptic
rings emerging from the magnetic ordering vector $(\pi,0)$.
This is not surprising since only magnon excitations come into play below the energy gap of quadrupolar branch.
With increasing energy, the rings increase with size, see Figs.~\ref{fig:DSFscan}(b) and (e).
The peculiar feature of the quadrupolar DSF is the presence of two concentric rings when the cutting energies are
above the gap. The inner one is attributed to the coherent excitations of quadrupolar mode.
At sufficiently high energy, the spectral weight of the dipolar DSF decreases greatly and ultimately disappears
as the cutting energy exceeds the band width of the magnon dispersion, see Fig.~\ref{fig:DSFscan}(c).
Thus in the energy interval between the maximum energy of magnon and quadrupolar dispersion,
only quadrupolar DSF persists which forms stripe patterns, see Fig.~\ref{fig:DSFscan}(f).

%%%%%%%%%%%%%%%%%%%%%%%%%%%%%%%%%%%%%%%%%%%%%%%%%%%%%%%%%%%%%%%%
\section{Summary and Conclusion}\label{Sec.conclusion}
%%%%%%%%%%%%%%%%%%%%%%%%%%%%%%%%%%%%%%%%%%%%%%%%%%%%%%%%%%%%%%%%
To summarize, we have studied the variational phase diagram of the extended BBQ model
by incorporating both NN and NNN exchange couplings on the square lattice.
Apart from the $(\pi,0)$ CAFM phase relevant to iron pnictides, various magnetic orderings
including conventional spin dipolar orders, novel quadrupolar orders
(spin nematic) and mixed dipolar-quadrupolar orders are identified.
This suggests that it is possible to find these novel orders in close proximity to the CAFM phase
in the phase diagram in other iron-based SCs, e.g. the stoichiometric FeSe~\cite{RelatedFeSe}.

We have also calculated the elementary excitation spectra in the $(\pi,0)$ CAFM phase
within the framework of SU(3) flavor-wave theory.
By fitting the experimental spin-wave dispersion, we have obtained the most relevant exchange constants for iron pnictides.
It is suggested that the NN bilinear coupling $J_1$ deduced from the SU(3) flavor-wave theory
differs strongly with the previous predictions by the conventional SU(2) spin-wave theory.

Finally, we have presented the dynamical correlations of both spin dipolar and quadrupolar components
for iron pnictides. The spin dipolar and quadrupolar DSFs can be directly probed with future experiments
in INS and optical spectroscopies, respectively.
Our results are consistent with and go beyond prior studies with the conventional SU(2) spin-wave
theory where the spin quadrupolar nature can not be captured.

\begin{acknowledgments}
C.L. and D.X.Y. acknowledge support from National Basic Research Program of China (Grant No. 2012CB821400), National Natural Science Foundation of China (Grant Nos. 11574404, 11275279), Natural Science Foundation of Guangdong Province (Grant No. 2015A030313176), Special Program for Applied Research on Super Computation of the NSFC-Guangdong Joint Fund, Guangdong Province Key Laboratory of Computational Science and the Guangdong Province Computational Science Innovative Research Team, Beijing Computational Science Research Center, and Fundamental Research Funds for the Central Universities of China. T.D. acknowledges Cottrell Research Corporation and Augusta University Scholarly Activity Award.
\end{acknowledgments}

%%%%%%%%%%%%%%%%%%%%%%%%%%%%%%%%%%%%%%%%%%%%%%%%%%%%%%%%%%%%%%%%
\appendix

%%%%%%%%%%%%%%%%%%%%%%%%%%%%%%%%%%%%%%%%%%%%%%%%%%%%%%%%%%%%%%%%
\section{SU(2) spin-wave theory for the BBQ model}\label{App:SU(2)SWT}
%%%%%%%%%%%%%%%%%%%%%%%%%%%%%%%%%%%%%%%%%%%%%%%%%%%%%%%%%%%%%%%%
Here we use the conventional SU(2) spin-wave theory to study the BBQ model (\ref{BBQmodel}).
The Schwinger representation of SU(2) algebra is defined by introducing two boson, $a$ and $b$.
The spin operators can be written as
\begin{align}
  S^+=a^\dag b,\ S^-=b^\dag a,\  S^z=\frac{1}{2}(a^\dag a-b^\dag b),
\end{align}
along with the constraint
\begin{align}\label{su2constraint}
  a^\dag a+b^\dag b=2S.
\end{align}
The Holstein-Primakoff (HP) transformation is introduced to desccribe the broken symmetry phases
by condensing one of the two bosons with the constraint (\ref{su2constraint}).
In the local rotating frame, we can define the following HP transformation
\begin{align}
 S_i^z=S-a_i^\dag a_i,\ S_i^-=a^\dag\sqrt{2S-a_i^\dag a_i},\ S_i^+=(S_i^-)^\dag.
\end{align}
The quadratic SU(2) spin-wave theory Hamiltonian of the BBQ model (\ref{BBQmodel}) in the $(\pi,0)$ CAFM phase is given by
\begin{align}
 \mathcal{H}=\sum_{\kk}A_{\kk}a_{\kk}^\dag a_{\kk}
 -\frac{B_{\kk}}{2}(a_{\kk}a_{-\kk}+\mathrm{H.c.}),
\end{align}
with
\begin{align}
  A_{\kk}=&4J_2+8(K_1+K_2)+(2J_1-4K_1)\cos k_y,\\
  B_{\kk}=&(2J_1+4K_1)\cos k_x+(4J_2+8K_2)\cos k_x\cos k_y.
\end{align}
It is worth noting that the above linear spin-wave Hamiltonian can be also obtained
from an effective $\tilde{J}_x-\tilde{J}_y-\tilde{J}_2$ model. Actually in the harmonic level the two models are
exactly related via the following Hubbard-Stratonovich
transformation~\cite{NatPhys.7.485,PhysRevB.85.144403,PhysRevB.86.085148,PhysRevB.92.165102}
\begin{align}
  \tilde{J}_x=J_1+2K_1,\ \tilde{J}_y=J_1-2K_1,\ \tilde{J}_2=J_2+2K_2,
\end{align}
which can be easily deduced by a mean-field decoupling of a pair of NN spins for the biquadratic term
\begin{align}
 (\bs_i\cdot\bs_j)^2\approx2\langle\bs_i\cdot\bs_j\rangle\bs_i\cdot\bs_j-\langle\bs_i\cdot\bs_j\rangle^2.
\end{align}
The resulting Hamiltonian can be diagonalized by a Bogoliubov transformation, leading to the elementary excitation spectrum
\begin{align}
 \varepsilon_{\kk}=\sqrt{A_{\kk}^2-B_{\kk}^2}.
\end{align}
The spin-wave dispersion for iron pnictides is shown in Fig.~\ref{fig:DispersionFit}
with the accepted parameters $J_1=22$ meV, $J_2=19$ meV and $K_1=14$ meV.

The dynamic spin correlations can be also calculated by expanding the HP transformation of spin operators to quadratic terms.
We find that the transverse DSF consists of only one type of excitations, namely the singe-magnon excitations
\begin{align}
\mathcal{S}^{T}(\qq,\omega)&=\frac{A_\kk-B_\kk}{\varepsilon_\kk}\delta(\omega-\varepsilon_{\kk}),
\end{align}
which shows a clear difference when compared with Eq.~(\ref{TS-DSF}) given by SU(3) flavor-wave theory.

\bibliographystyle{apsrev4-1}
\bibliography{RefBBQ}

\end{document}